\documentclass[final,5p,times,twocolumn]{elsarticle}

\graphicspath{{tutu/}}

\usepackage{amsmath}
\usepackage{amssymb}
\usepackage[T1]{fontenc}
\usepackage{placeins}

\journal{Nuclear Physics B}
\biboptions{numbers,sort&compress}

\begin{document}

\begin{frontmatter}

\title{Impact of Evaporation Barriers on Solar-Captured Dark Matter Distribution and Evaporation Mass}

\author[inst1,inst2,inst3]{Xuan Wen}
\ead{wenxuan23@mails.ucas.ac.cn}

\affiliation[inst1]{organization={School of Fundamental Physics and Mathematical Sciences, Hangzhou Institute for Advanced Study, UCAS}, city={Hangzhou}, postcode={310024}, country={China}}
\affiliation[inst2]{organization={Institute of Theoretical Physics, Chinese Academy of Sciences}, city={Beijing}, postcode={100190}, country={China}}
\affiliation[inst3]{organization={University of Chinese Academy of Sciences (UCAS)}, city={Beijing}, postcode={100049}, country={China}}

\begin{abstract}
Evaporation determines the low-mass reach of solar-captured dark matter because that reach is controlled by the small population of particles closest to the escape threshold. We present an orbit-space calculation of the non-thermal distribution of captured dark matter in the presence of an evaporation barrier generated by a smooth in-medium attraction sourced by the solar medium. We show that the barrier not only deepens the effective potential but also reshapes the near-threshold phase-space structure, displacing the equilibrium distribution away from weakly bound, escape-prone trajectories and toward more tightly bound core-crossing orbits, thereby suppressing evaporation and lowering the evaporation mass. Although the bulk population remains near thermal equilibrium, the near-threshold tail, as reflected in the projected velocity spectrum, acquires characteristic non-thermal structure because the barrier deforms the bound orbit space and preferentially retains particles that repeatedly traverse the hot solar core. The near-threshold tail is therefore essential for determining the low-mass reach of solar dark-matter searches in the barrier regime, and our orbit-space treatment captures the relevant physics in a controlled way.
\end{abstract}


\end{frontmatter}

\section{Introduction}\label{introduction}

The nature of dark matter (DM) remains a central open question in particle physics and cosmology, motivating a broad experimental program spanning underground direct searches, collider production, and indirect searches using astrophysical messengers~\cite{Jungman:1995df,Gaskins:2016cha,Lavalle:2012ef,LHAASO:2022yxw}. In the GeV--TeV mass regime, nuclear-recoil searches have achieved high sensitivity. Multi-tonne liquid-xenon detectors place stringent constraints on DM--nucleon scattering across a broad range of DM masses~\cite{LZ:2022lsv,XENON:2023cxc,PandaX:2024qfu}. For sub-GeV DM, the typical nuclear-recoil energies fall below conventional thresholds, motivating low-threshold detector concepts and alternative channels that extend sensitivity to light DM~\cite{Essig:2012yx,Crisler:2018gci,SuperCDMS:2022kse}.

DM capture in celestial bodies offers a complementary probe of DM interactions~\cite{Press:1985ug,Gould:1987ir}. In the Sun, halo DM can scatter off solar nuclei and lose sufficient kinetic energy to become gravitationally bound, leading to the accumulation of DM in the solar interior~\cite{Gould:1987ju}. The subsequent evolution of the captured population is shaped by capture, annihilation, and evaporation, and can be characterized by the corresponding equilibration timescales~\cite{Liang:2016yjf}. If captured DM annihilates, the annihilation products can yield high-energy neutrinos that escape the Sun and can be probed by existing neutrino telescopes such as Super-Kamiokande, IceCube, and ANTARES~\cite{Super-Kamiokande:2015xms,IceCube:2021xzo,ANTARES:2016xuh}.
The sensitivity to such signals will be further extended by next-generation facilities, including IceCube-Gen2, KM3NeT, and TRIDENT~\cite{IceCube-Gen2:2020qha,Adrian-Martinez:2016fdl,TRIDENT:2022hql}.

The neutrino signal from the Sun is governed by the competition among capture, annihilation, and evaporation. For light DM, evaporation can substantially deplete the solar population and suppress the annihilation signal, so the low-mass reach of solar DM searches is controlled by the phase-space population closest to the escape threshold.

Recent work has shown that additional long-range DM--Standard Model (SM) interactions mediated by a light particle can modify the evaporation mass of solar-captured DM.
In particular, Ref.~\cite{Acevedo:2023owd} demonstrated that, when the mediator range exceeds microscopic length scales of the solar medium, the SM density sources a smooth in-medium potential that creates an evaporation barrier.
This barrier deepens the effective potential well and raises the escape threshold, suppressing evaporation and shifting the evaporation mass to lower values.
In that treatment, the captured population is described by the radial number-density profile $n_\chi(r)$, obtained from standard approximations for the bulk population in the collisional Boltzmann equation for contact DM--SM scattering~\cite{Spergel:1985kj,Gould:1987ju,Gould:1990thc,Gould:1990knu}. The short-mean-free-path regime is modeled in local thermal equilibrium, while the long-mean-free-path regime is approximated by an isothermal profile~\cite{Spergel:1985kj,Gould:1987ju,Gould:1990thc,Gould:1990knu}. The mediator-induced effect is then incorporated as an additional interaction potential contributing to the total potential relevant for DM motion~\cite{Acevedo:2023owd}.
While these approximations describe the bulk population well, evaporation is driven by rare energetic up-scattering events that populate the extreme tail near the escape boundary, where the phase-space structure of weakly bound orbits becomes essential~\cite{Gould:1987ju,Liang:2016yjf}.

In this work, we study how a prescribed SM-sourced evaporation barrier reshapes the non-thermal distribution of captured DM and shifts the evaporation mass. We determine this non-thermal distribution using the numerical orbit-space approach of Ref.~\cite{Liang:2016yjf}, which resolves the near-threshold tail beyond standard thermal approximations. Throughout, we treat the barrier as a fixed background contribution to the total potential and isolate its transport consequences for the bound distribution and the evaporation mass. The detailed interaction assumptions and the adopted potential are specified in Sec.~\ref{sec:distribution}.

\section{Framework and orbit-space distribution of captured dark matter}
\label{sec:distribution}

Standard bulk approximations for the captured population, such as local thermal equilibrium in the optically thick regime and an isothermal profile in the opposite limit, describe the bulk distribution well~\cite{Gould:1987ju,Gould:1990thc,Gould:1990knu}. Evaporation, however, is governed by the extreme tail of the bound population adjacent to the escape threshold and is fed by rare up-scattering events in the hot, dense core. In this near-threshold regime the occupation depends strongly on angular momentum, which determines how efficiently trajectories probe the core, and on the radial profile of the escape condition once the barrier is included. Because the evaporation rate depends exponentially on this tail, resolving its phase-space structure is essential for a controlled description of evaporation once the barrier reshapes the escape boundary~\cite{Busoni:2013kaa,Kouvaris:2015nsa}.

\subsection{Potential and scattering framework}

We consider an additional attractive interaction between DM and SM particles mediated by a light real scalar field $\varphi$ of mass $m_\phi$, with couplings $g_\chi$ to DM and $g_{\rm SM}$ to the relevant SM source. In a medium with a number-density profile of SM particles $n_{\rm SM}(\mathbf r)$, the corresponding static in-medium potential energy experienced by a DM particle takes the Yukawa form~\cite{Acevedo:2023owd,Knapen:2017xzo}
\begin{equation}
\phi_{\rm barrier}(\mathbf{r})
= -\,g_{\rm SM}g_\chi
\!\int_V\!
\frac{n_{\rm SM}(\mathbf{r'})\,e^{-m_\phi |\mathbf{r-r'}|}}
{4\pi |\mathbf{r-r'}|}\,d^3\mathbf{r'}\,.
\label{eq:phi_barrier_int}
\end{equation}
The evaporation barrier regime corresponds to mediator ranges $r_Y=m_\phi^{-1}$ that exceed microscopic length scales of the solar medium such as the interparticle spacing, so that the SM constituents act as a smooth source. When the SM density varies slowly within the interaction volume defined by the mediator range, Eq.~\eqref{eq:phi_barrier_int} reduces to the local-density form
\begin{equation}
\phi_{\rm barrier}(r)\simeq
-\,\frac{g_{\rm SM}g_\chi}{m_\phi^{2}}\,n_{\rm SM}(r)\,.
\label{eq:phi_barrier}
\end{equation}
For mediator ranges comparable to the solar radius, finite-size effects become important and the full Yukawa expression in Eq.~\eqref{eq:phi_barrier_int} should be retained~\cite{Acevedo:2023owd}.

We parametrize the central strength of the additional potential by
\begin{equation}
\beta \equiv \frac{|\phi_{\rm barrier}(0)|}{|\phi_{\rm grav}(0)|},
\label{eq:beta_def}
\end{equation}
which measures the scalar-induced contribution relative to gravity at the solar center. Since different choices of the mediator mass and couplings can lead to the same value of $\beta$, we use $\beta$ as a convenient phenomenological parameter for the depth of the potential well, which in turn controls the evaporation mass. In the following, $\beta=0$ denotes the gravity-only case, while $\beta=1,2,3$ label the representative barrier benchmarks studied in detail.

Parameter values with $\beta=\mathcal{O}(1)$ arise in explicit ultralight-mediator realizations of an evaporation barrier, as illustrated in Appendix~C of Ref.~\cite{Acevedo:2023owd}. In the present work, however, $\beta$ is treated purely as a phenomenological parameter specifying the depth of a prescribed SM-sourced barrier, rather than as a complete specification of the underlying mediator model. In a minimal scalar Yukawa realization, the captured DM would in principle also source an additional mean field and hence backreact on the total potential. We do not include that DM-sourced contribution here, because our aim is to isolate the transport consequences of a prescribed SM-sourced evaporation barrier for the non-thermal distribution and evaporation. Accounting for this feedback would require solving simultaneously for the mediator profile and the corresponding DM non-thermal distribution in the resulting total potential, which is beyond the scope of the present work.

Capture and thermalization are described by the simplest spin-independent interaction in the non-relativistic effective theory, represented by the operator $\hat{\mathcal O}_1=\mathbf{1}$ and parametrized by a reference DM--proton cross section $\sigma_p$~\cite{Fan:2010gt,Fitzpatrick:2012ix,Anand:2013yka,Liang:2016yjf,Widmark:2017yvd}. In practice, the same $\hat{\mathcal O}_1$ interaction is used throughout for capture, thermalization, and the orbit-to-orbit scattering rates entering the Boltzmann evolution. The light mediator is included only through an additional smooth in-medium potential that modifies binding and escape conditions. This factorized treatment applies when mediator-induced contributions to the binary DM--SM differential cross section remain subdominant at the momentum transfers relevant to solar capture and thermalization, while the same mediator still generates a non-negligible coherent mean field in the solar medium, as discussed in Ref.~\cite{Acevedo:2023owd}. This separation is justified because the barrier is sourced by the SM density of the solar medium in the near-static, long-range limit, whereas capture and thermalization are controlled by finite-momentum-transfer scattering processes.

As a consistency check of our implementation, we have verified that, in the gravity-only limit $\beta = 0$, our orbit-space solution is consistent with the non-thermal distribution and evaporation behavior obtained in Ref.~\cite{Liang:2016yjf} under the same solar model and scattering assumptions. This supports interpreting the modifications discussed below as consequences of the in-medium potential.

\subsection{Non-thermal distribution over bound orbits}

We describe the captured population by an orbit-space distribution $f_\chi(E,L)$ over bound trajectories labeled by the specific energy $E$ and angular momentum $L$ in the adopted potential. Collisions with solar nuclei induce transitions among these orbital integrals, leading to a linear collisional Boltzmann equation restricted to the bound region~\cite{Liang:2016yjf}. The corresponding orbit-to-orbit transition rates are obtained by integrating the local collision rate along representative trajectories and averaging over the thermal motion of the target nuclei, while scatterings that map the post-collision orbit outside the bound region are counted as evaporation~\cite{Garani:2017jcj}. Evolving the system to late times yields the non-thermal distribution over bound orbits, from which the configuration-space and velocity distributions follow by averaging along each orbit.

We adopt the fixed, spherically symmetric background potential defined above and treat the motion of captured particles between scatterings as collisionless. The in-medium contribution is asymptotically negligible at large radii. Inside the Sun, it deepens the potential well and therefore increases the local escape velocity relative to the gravity-only scenario.

Neglecting DM self-interactions, we average the collisional Boltzmann equation over bound orbits to obtain a closed master equation for $f_\chi(E,L)$ on the bound region of $(E,L)$ space.
After discretizing that domain, the evolution takes the form
\begin{equation}
\begin{aligned}
\frac{d f_\chi(E,L)}{dt}
= {}& -\,f_\chi(E,L)\!\!\sum_{E',L'} S(E,L;E',L')
\\
& + \!\!\sum_{E',L'} f_\chi(E',L')\,S(E',L';E,L)\,,
\end{aligned}
\label{eq:EL-master}
\end{equation}
where $S(E,L;E',L')$ is the orbit-averaged scattering rate for a single collision that maps $(E,L)$ to $(E',L')$.
Transitions that place the post-collision trajectory outside the bound region defined by $\phi_{\rm tot}(r)$ are treated separately when computing evaporation. Here $f_\chi$ is treated as a unit-normalized equilibrium shape distribution over the bound region, while the absolute abundance is tracked separately by the total captured number $N_\chi$ obtained from the global population evolution.

We evaluate $S(E,L;E',L')$ using the weighted Markov chain Monte Carlo (MCMC) orbit method of Ref.~\cite{Liang:2016yjf}.
For each initial label $(E,L)$ we integrate the corresponding trajectory in the adopted potential and calculate the orbit-averaged rate by sampling the local collision probability along the orbit.
At each sampled position, we select a nuclear species according to its partial rate, draw the target velocity from the local Maxwell-Boltzmann distribution, evaluate the scattering kinematics in the center-of-mass frame for the assumed interaction, and transform back to the solar frame to obtain $(E',L')$.
Accumulating the weighted contributions over the orbit yields the entries of the transition matrix $S$.

To obtain the non-thermal distribution $f_\chi(E,L)$, we evolve Eq.~\eqref{eq:EL-master} on the discretized bound region using the MCMC estimate of $S(E,L;E',L')$.
In optically thin or kinematically suppressed regimes, relaxation to equilibrium can be slow, leaving a residual dependence on the capture and thermalization history if instantaneous equilibration is not imposed~\cite{Widmark:2017yvd}.
We quantify this dependence by repeating the evolution from three representative classes of initial profiles on the $(E,L)$ domain, chosen to bracket plausible capture and thermalization histories. Specifically, we consider an initial distribution biased toward high-energy orbits near the escape boundary, one concentrated in the most tightly bound region, and one with broad support across the full bound domain.
The resulting spread is reported as the shaded band in the distributions below.

\begin{figure*}[!t]
  \centering
  \setlength{\tabcolsep}{3pt}
  \renewcommand{\arraystretch}{1.0}
  \begin{tabular}{cc}
    \includegraphics[width=0.47\textwidth]{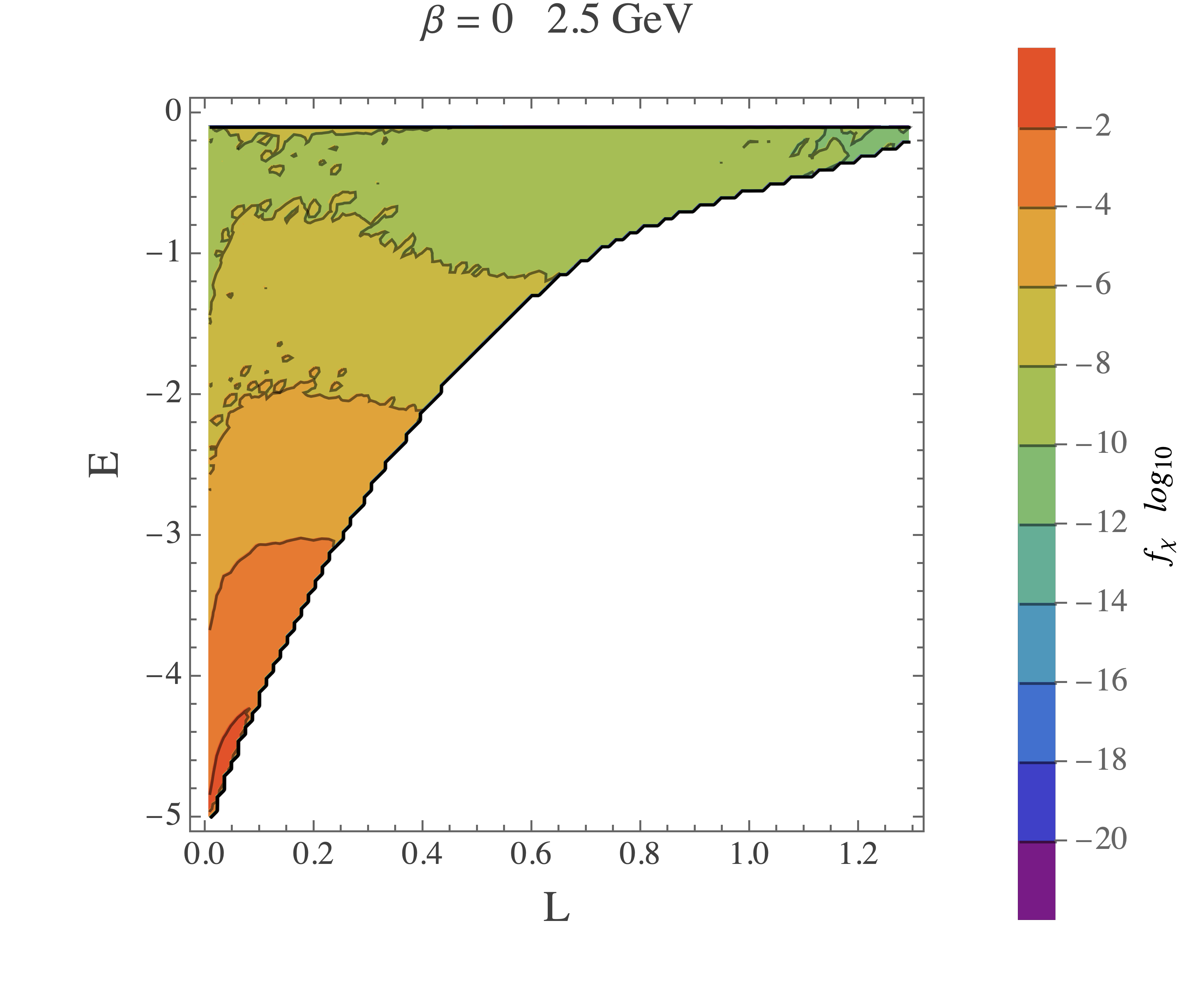} &
    \includegraphics[width=0.47\textwidth]{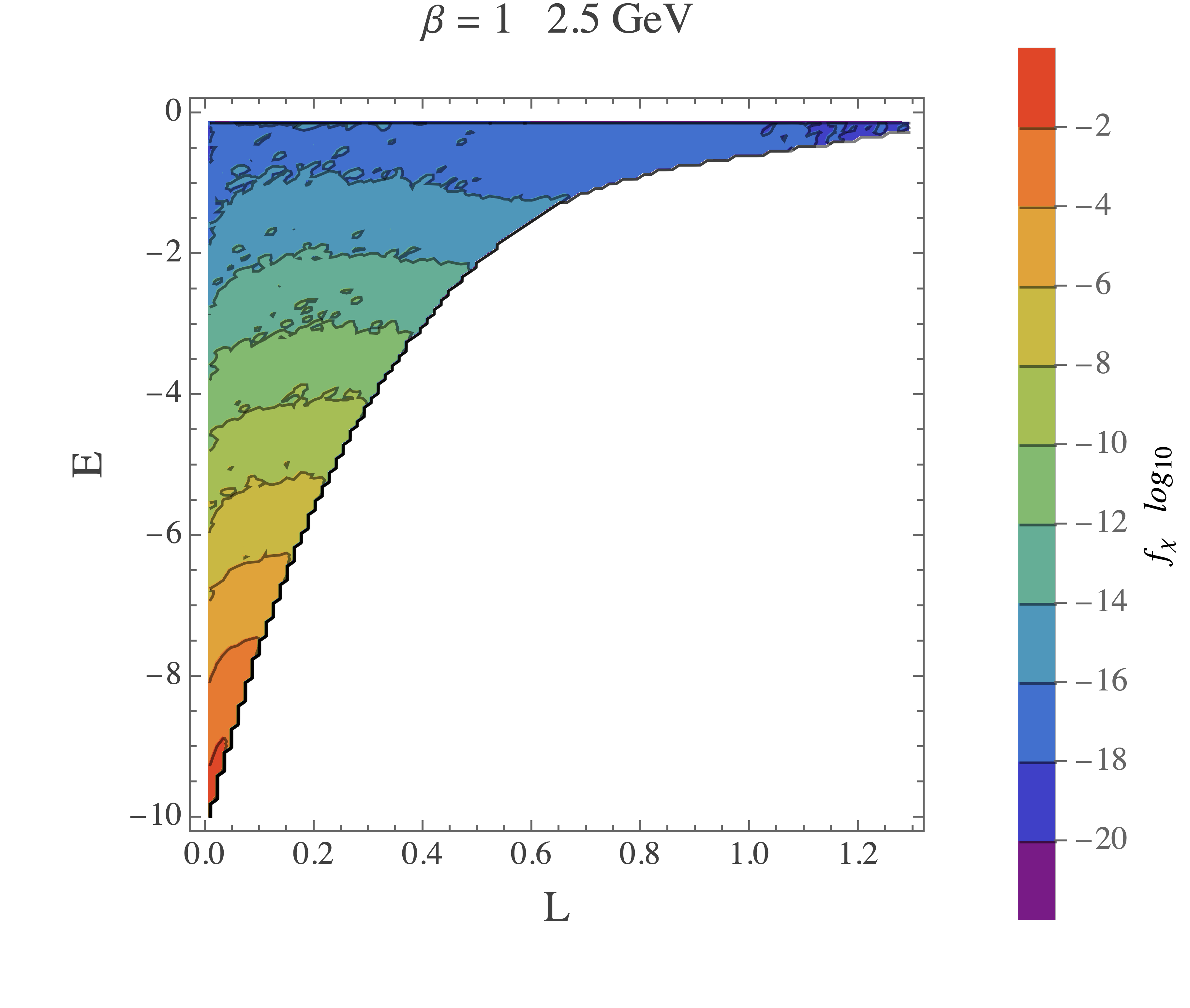}
  \end{tabular}
  \caption{Non-thermal orbit-space distribution $f_\chi(E,L)$ for $m_\chi=2.5\,\mathrm{GeV}$ in the cases $\beta=0$ (left) and $\beta=1$ (right). The energy $E$ and angular momentum $L$ are shown in units of $GM_\odot/R_\odot$ and $\sqrt{GM_\odot R_\odot}$, respectively. The shaded regions denote the domains of bound orbits.}

  \label{fig:ELcontours}
\end{figure*}

Figure~\ref{fig:ELcontours} shows the non-thermal orbit distribution $f_\chi(E,L)$ for $m_\chi=2.5\,\mathrm{GeV}$ in the cases $\beta=0$ and $\beta=1$.
Including the in-medium potential deepens the interior well and raises $v_{\rm esc}(r)$, which deforms the bound region in the $(E,L)$ plane.
The occupation shifts toward more tightly bound orbits and smaller angular-momentum values.
The preference for small $L$ reflects the tendency of more radial trajectories to probe the hot, dense core where most scatterings occur.
The deformation is most pronounced near the escape boundary, where the small population of orbits that can be promoted above the escape threshold by a single energetic scattering resides~\cite{Kouvaris:2015nsa}.

\subsection{From orbit space to local phase space}

For presentation purposes, the orbit-space and local phase-space variables shown below are reported in dimensionless form. Specifically, the plotted variables are $E/(GM_\odot/R_\odot)$, $L/\sqrt{GM_\odot R_\odot}$, $r/R_\odot$, and $v/\sqrt{GM_\odot/R_\odot}$, with $R_\odot=6.955\times10^5~\mathrm{km}$. The associated reference scales are $(GM_\odot/R_\odot^3)^{-1/2}\simeq 1.596\times 10^3~\mathrm{s}$ and $\sqrt{GM_\odot/R_\odot}\simeq 436~\mathrm{km}\,\mathrm{s}^{-1}$. Unless stated otherwise, the orbit-space, local phase-space, and projected velocity distributions shown below use these dimensionless variables. Thus a velocity axis labeled by $v_\chi$ should be read as $v_\chi/\sqrt{GM_\odot/R_\odot}$.

\begin{figure*}[!t]
\centering
\setlength{\tabcolsep}{2pt}
\renewcommand{\arraystretch}{1.0}

\begin{tabular}{cc}
\includegraphics[width=0.385\textwidth]{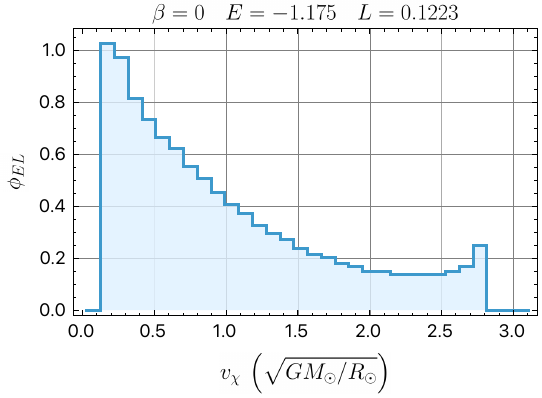} &
\includegraphics[width=0.385\textwidth]{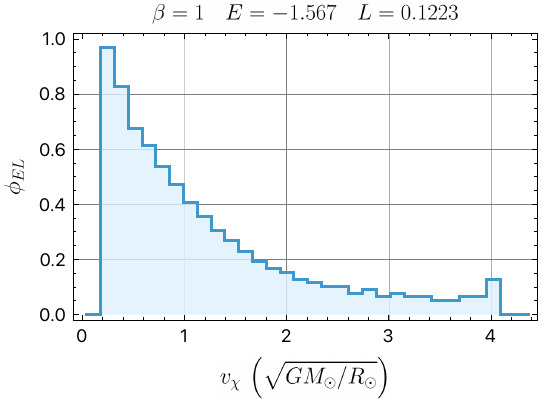} \\[1pt]
\includegraphics[width=0.385\textwidth]{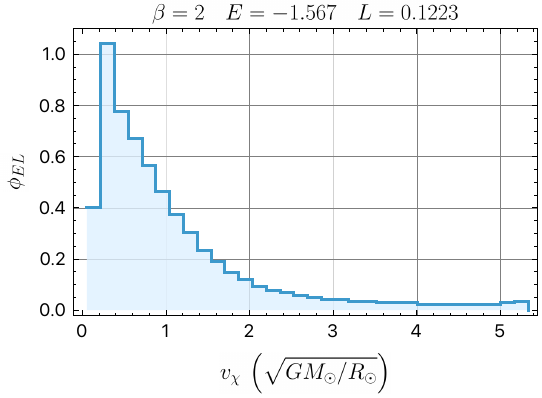} &
\includegraphics[width=0.385\textwidth]{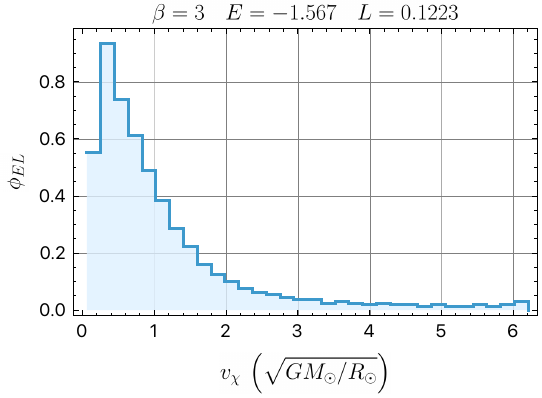}
\end{tabular}

\caption{Velocity marginal distributions of the residence-time weight $\phi_{EL}$ for an illustrative bound orbit. The panels, ordered from top left to bottom right, correspond to $\beta=0,1,2,3$ at fixed angular momentum $L=0.1223$. The specific energy is $E=-1.175$ for $\beta=0$ and $E=-1.567$ for $\beta=1,2,3$.}
\label{fig:XVD-velocity}

\vspace{1pt}

\begin{tabular}{cc}
\includegraphics[width=0.385\textwidth]{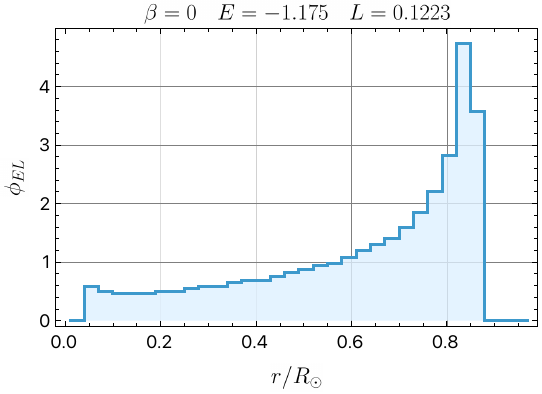} &
\includegraphics[width=0.385\textwidth]{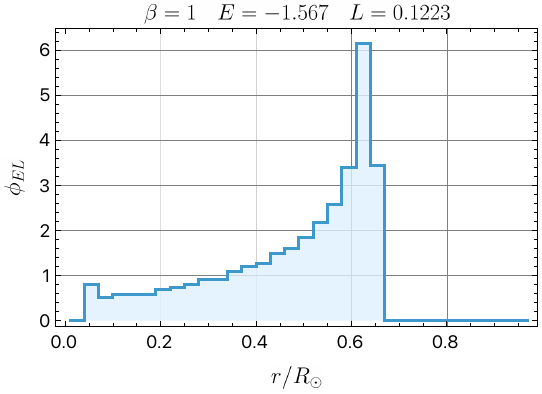} \\[1pt]
\includegraphics[width=0.385\textwidth]{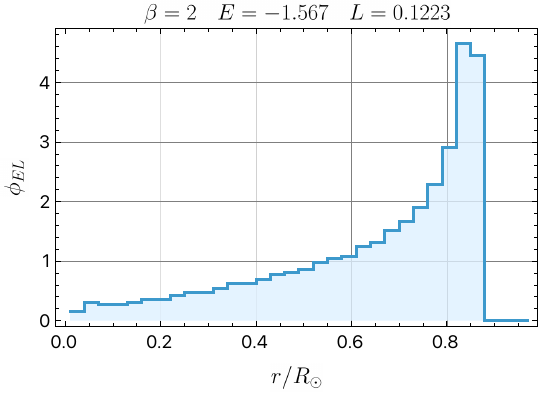} &
\includegraphics[width=0.385\textwidth]{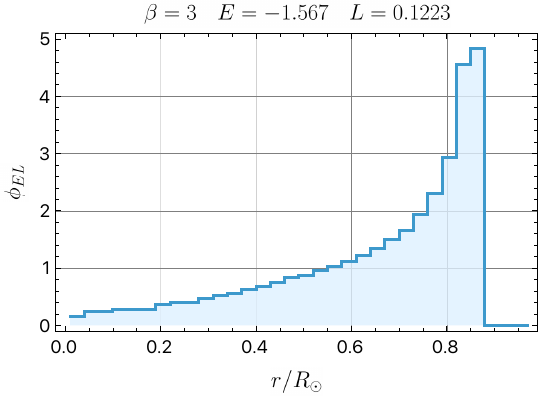}
\end{tabular}

\caption{Radial marginal distributions of the residence-time weight $\phi_{EL}$ for the same illustrative orbit as in Fig.~\ref{fig:XVD-velocity}. The panels, ordered from top left to bottom right, correspond to $\beta=0,1,2,3$.}
\label{fig:XVD-radial}
\end{figure*}

Starting from the non-thermal orbit-space distribution $f_\chi(E,L)$ over bound trajectories in the adopted potential, we construct the corresponding local distribution in terms of radius $r$ and velocity $v$ in the solar rest frame.
Since solar scattering occurs only when the trajectory intersects the Sun, the orbit sample entering the collision integrals is restricted to bound trajectories with perihelia inside $R_\odot$.
For each retained pair $(E,L)$ we define $\phi_{EL}(r,v)$ as the residence-time density over the full bound orbit, normalized over its full orbital support.
When evaluating solar scattering or evaporation, only the radial segment with $0\le r\le R_\odot$ is retained. The integral of $\phi_{EL}$ over this restricted domain equals the fraction of the orbital period spent inside the Sun.
For the solar-interior projections shown below, we discretize $0\le r\le R_\odot$ and $0\le v\le v_{\rm esc}(r)$, where $v_{\rm esc}(r)$ is the escape velocity in the adopted total potential.
The local $(r,v)$ distribution then follows as
\begin{equation}
f_\chi(r,v)=\sum_{E,L} f_\chi(E,L)\,\phi_{EL}(r,v)\,,
\label{eq:mapping}
\end{equation}
where the sum runs over the discretized bound region in $(E,L)$ introduced above. Figures~\ref{fig:XVD-velocity} and~\ref{fig:XVD-radial} illustrate, for a sample bound orbit, the velocity and radial marginal distributions of $\phi_{EL}$ obtained by integrating over $r$ and $v$, respectively.
This construction is purely dynamical and does not assume an isothermal or Maxwell--Boltzmann form for the captured population. It applies equally in the gravity-only and in-medium cases, with the latter modifying the kinematics only through the $\beta$-dependent total potential~\cite{Liang:2016yjf}.

\subsection{Thermal reference and non-thermal deviations}
For comparison, we introduce an isothermal reference distribution,
$f_{\rm th}(E)\propto \exp\!\left(-m_\chi E/T_\chi\right)$, characterized by a single effective temperature $T_\chi$~\cite{Gould:1987ju,Liang:2016yjf}.
In the long-mean-free-path limit, the captured population is often approximated as isothermal~\cite{Spergel:1985kj,Gould:1987ju,Gould:1990knu,Busoni:2017mhe}.
We fix $T_\chi$ by requiring that the net energy exchange with the solar medium vanishes once equilibrium is achieved~\cite{Gould:1987ju,Widmark:2017yvd}, which yields

\begin{equation}
\begin{aligned}
\sum_A \int_0^{R_\odot}\!dr\; r^2\, n_A(r)\,
&\left[\frac{m_A T_\chi + m_\chi T_\odot(r)}{m_A m_\chi}\right]^{1/2}
\\
&\times \big[T_\odot(r)-T_\chi\big]\,
\exp\!\left[-\,\frac{m_\chi\,\phi_{\rm tot}(r)}{T_\chi}\right]
= 0\,,
\end{aligned}
\label{eq:Tx_condition}
\end{equation}
where $n_A(r)$ is the number-density profile of nuclear species $A$, and $T_\odot(r)$ is the solar temperature profile.
Our scattering calculation includes the elements H, $^4$He, $^{14}$N, $^{16}$O, and $^{56}$Fe from the Standard Solar Model GS98~\cite{Grevesse:1998bj}.
In the discussion below we quantify deviations from isothermality using the ratio $f_\chi/f_{\rm th}$ of the projected velocity distributions.
This ratio isolates non-thermal structure and reduces sensitivity to normalization conventions.
Here the projected velocity distribution is obtained by integrating $f_\chi(r,v)$ over radius inside the Sun.

\begin{figure*}[!t]
\centering
\begin{minipage}[t]{0.49\textwidth}
\centering
\setlength{\tabcolsep}{1.5pt}
\renewcommand{\arraystretch}{1.0}
\begin{tabular}{cc}
\includegraphics[width=0.485\linewidth]{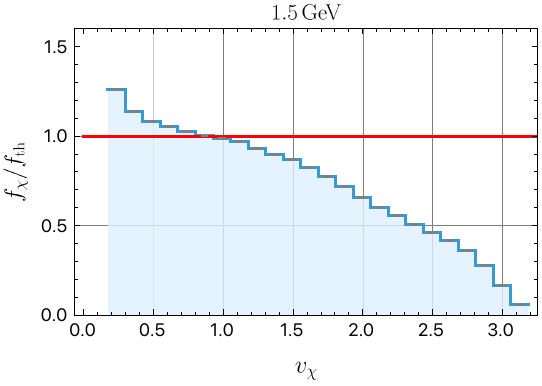} &
\includegraphics[width=0.485\linewidth]{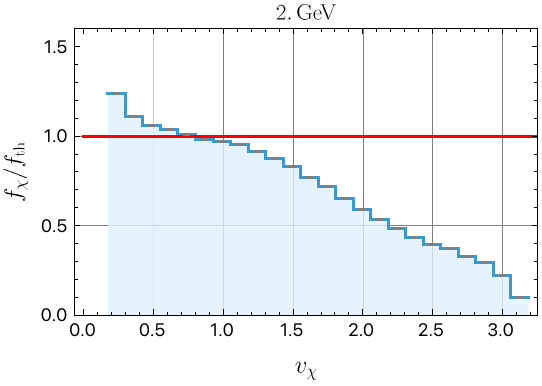} \\[1pt]
\includegraphics[width=0.485\linewidth]{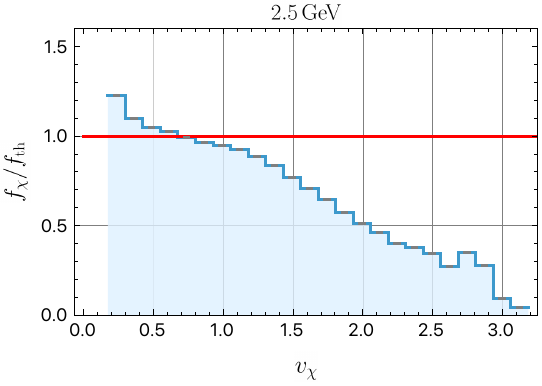} &
\includegraphics[width=0.485\linewidth]{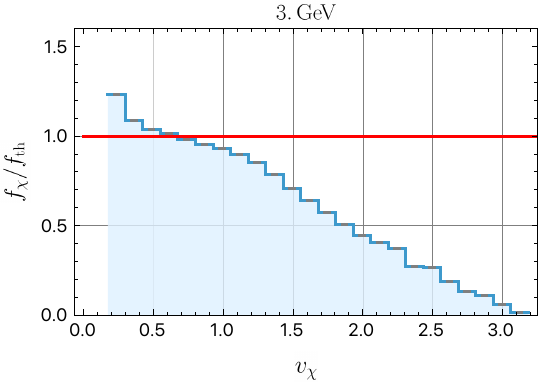} \\[1pt]
\includegraphics[width=0.485\linewidth]{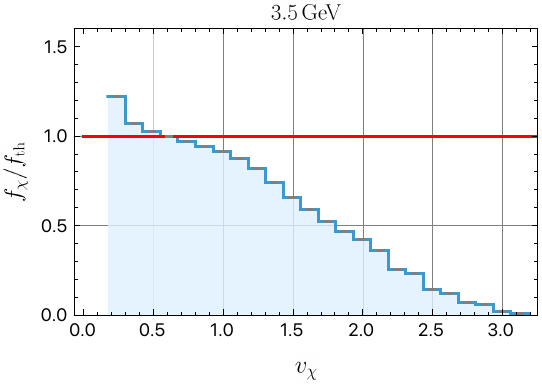} &
\includegraphics[width=0.485\linewidth]{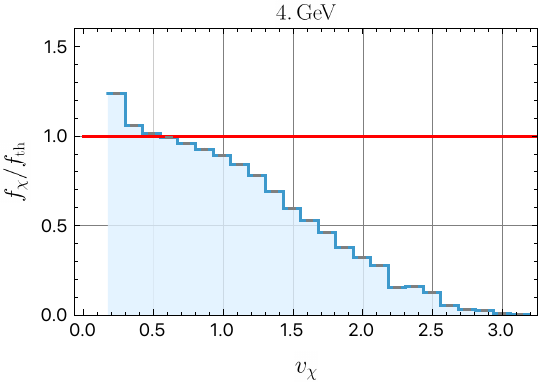}
\end{tabular}

\smallskip
{\small (a) $\beta=0$.}
\end{minipage}\hfill
\begin{minipage}[t]{0.49\textwidth}
\centering
\setlength{\tabcolsep}{1.5pt}
\renewcommand{\arraystretch}{1.0}
\begin{tabular}{cc}
\includegraphics[width=0.485\linewidth]{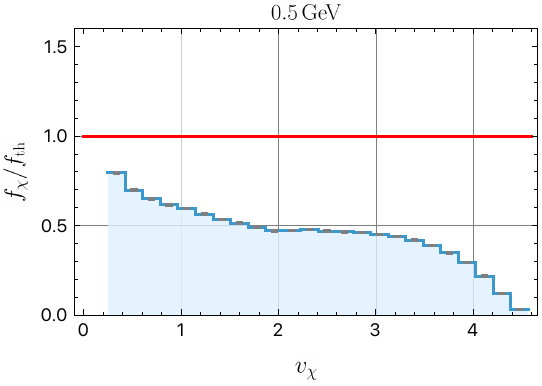} &
\includegraphics[width=0.485\linewidth]{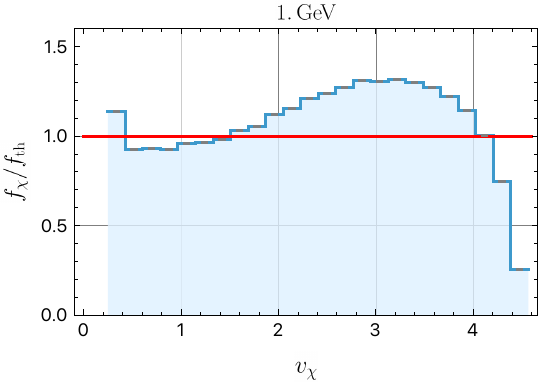} \\[1pt]
\includegraphics[width=0.485\linewidth]{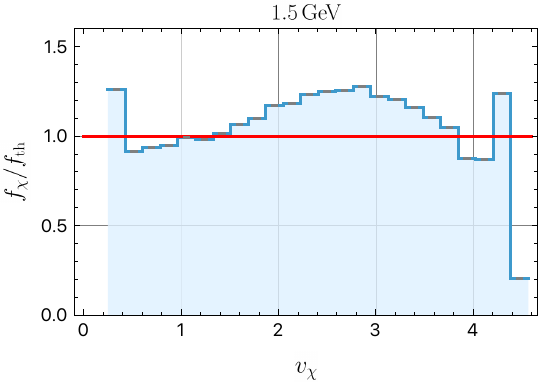} &
\includegraphics[width=0.485\linewidth]{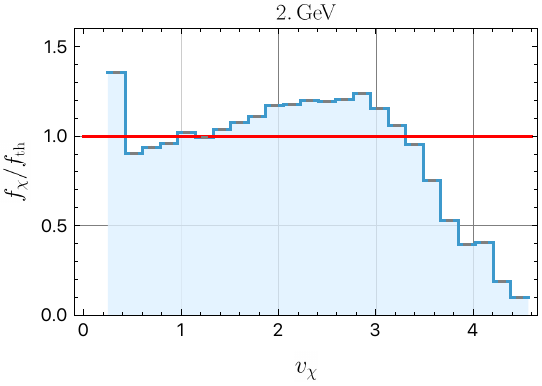} \\[1pt]
\includegraphics[width=0.485\linewidth]{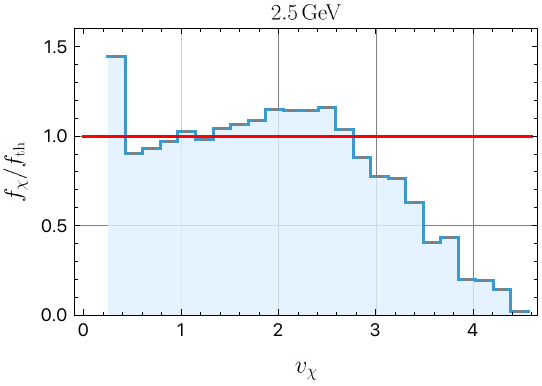} &
\includegraphics[width=0.485\linewidth]{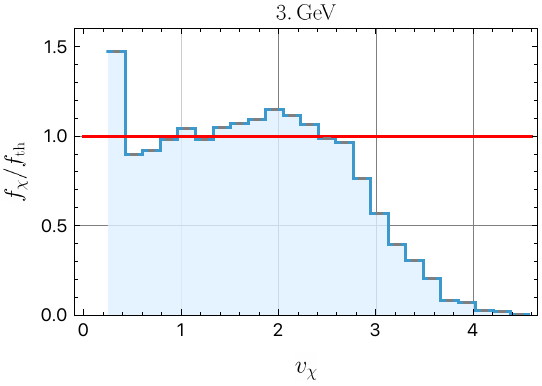} \\[1pt]
\includegraphics[width=0.485\linewidth]{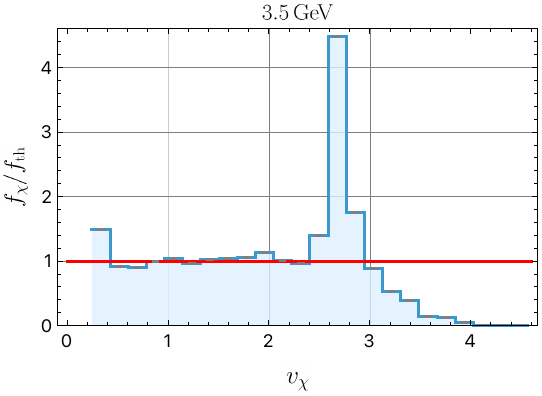} &
\includegraphics[width=0.485\linewidth]{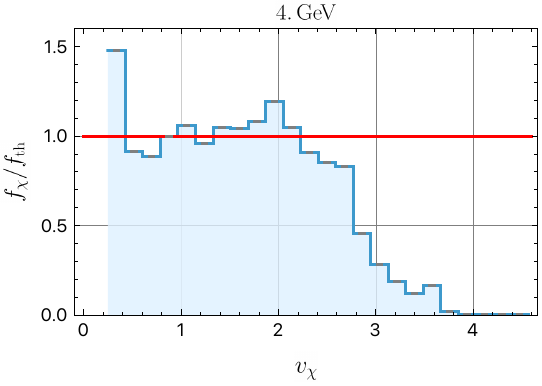}
\end{tabular}

\smallskip
{\small (b) $\beta=1$.}
\end{minipage}
\caption{Ratio $f_\chi/f_{\rm th}$ of the projected non-thermal velocity distribution to the isothermal reference for $\beta=0$ and $\beta=1$. Group (a) corresponds to $\beta=0$ and spans $m_\chi=1.5$--$4.0~\mathrm{GeV}$ in steps of $0.5~\mathrm{GeV}$. Group (b) corresponds to $\beta=1$ and spans $m_\chi=0.5$--$4.0~\mathrm{GeV}$ in steps of $0.5~\mathrm{GeV}$. The shaded band in group (a) shows the spread among the three initial profiles discussed in the text.}
\label{fig:ratio-betas01}
\end{figure*}

\begin{figure*}[!t]
\centering
\begin{minipage}[t]{0.49\textwidth}
\centering
\setlength{\tabcolsep}{1.5pt}
\renewcommand{\arraystretch}{1.0}
\begin{tabular}{cc}
\includegraphics[width=0.485\linewidth]{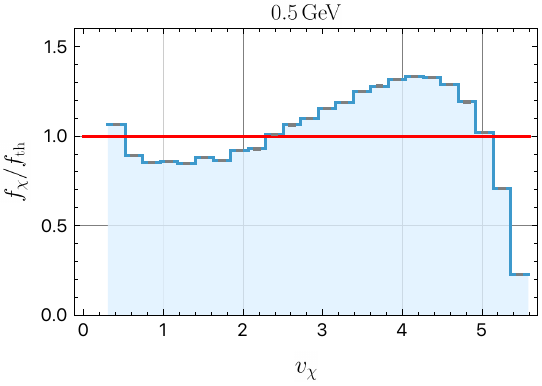} &
\includegraphics[width=0.485\linewidth]{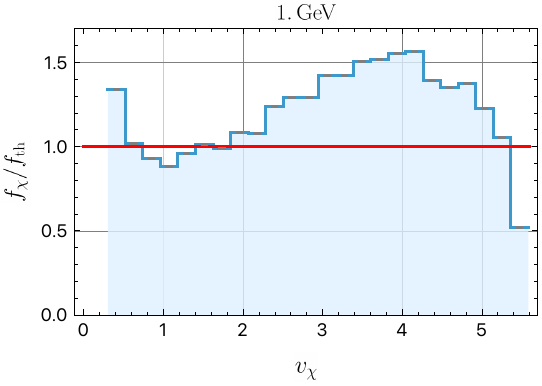} \\[1pt]
\includegraphics[width=0.485\linewidth]{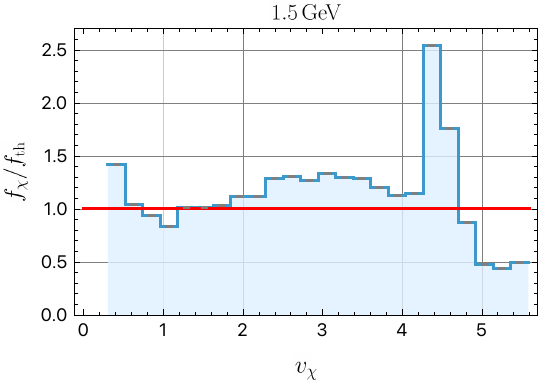} &
\includegraphics[width=0.485\linewidth]{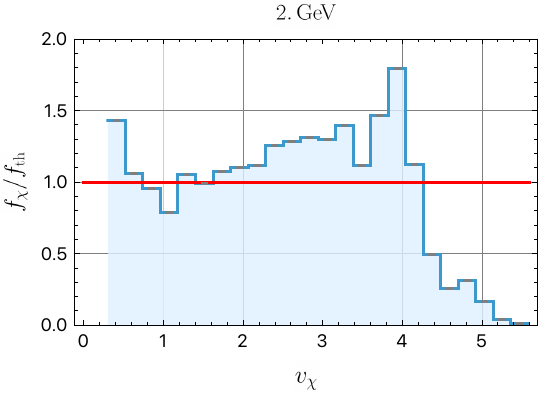} \\[1pt]
\includegraphics[width=0.485\linewidth]{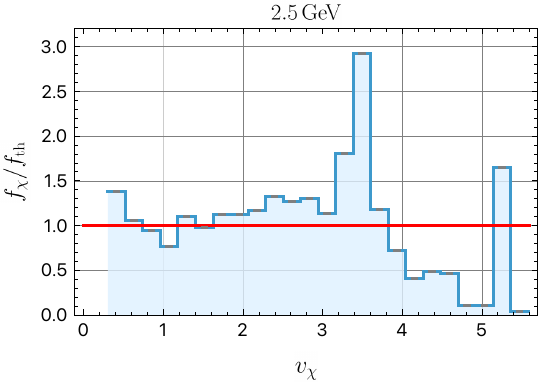} &
\includegraphics[width=0.485\linewidth]{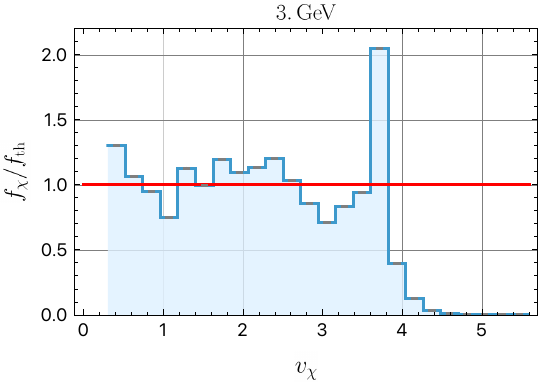} \\[1pt]
\includegraphics[width=0.485\linewidth]{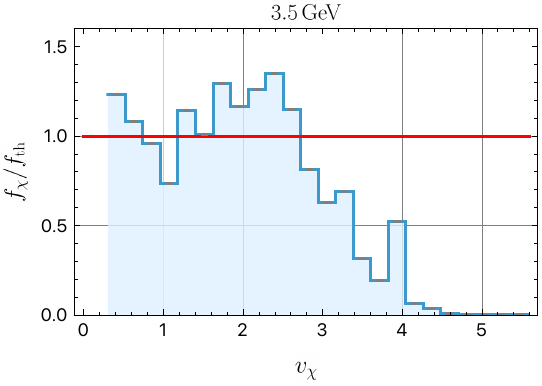} &
\includegraphics[width=0.485\linewidth]{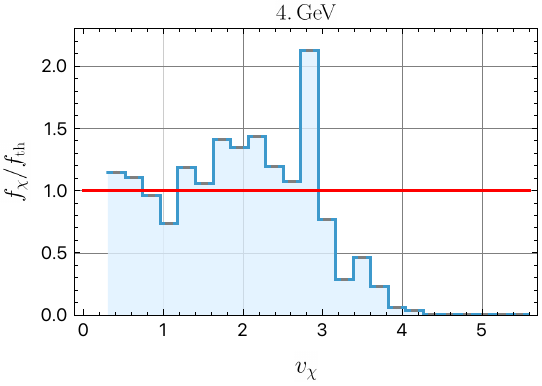}
\end{tabular}

\smallskip
{\small (a) $\beta=2$.}
\end{minipage}\hfill
\begin{minipage}[t]{0.49\textwidth}
\centering
\setlength{\tabcolsep}{1.5pt}
\renewcommand{\arraystretch}{1.0}
\begin{tabular}{cc}
\includegraphics[width=0.485\linewidth]{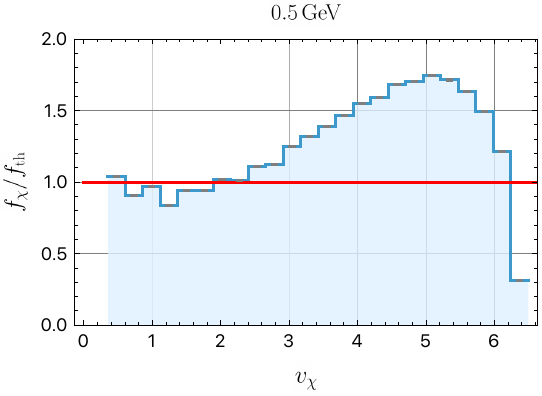} &
\includegraphics[width=0.485\linewidth]{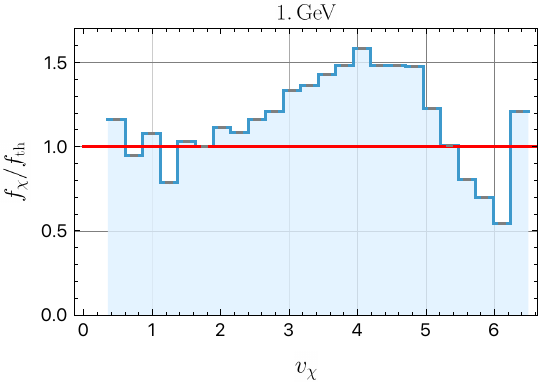} \\[1pt]
\includegraphics[width=0.485\linewidth]{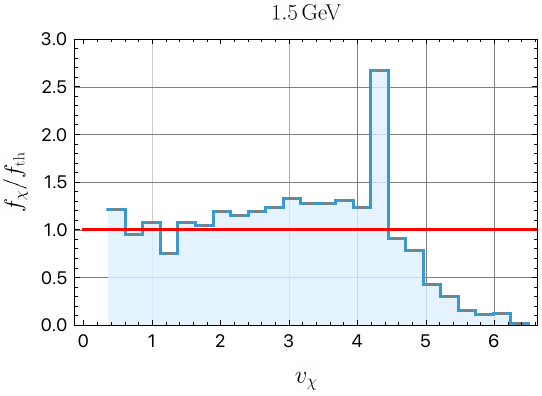} &
\includegraphics[width=0.485\linewidth]{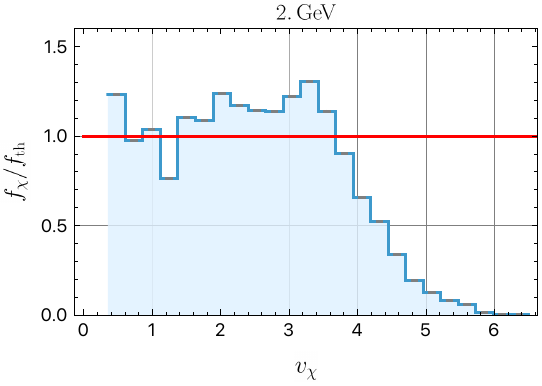} \\[1pt]
\includegraphics[width=0.485\linewidth]{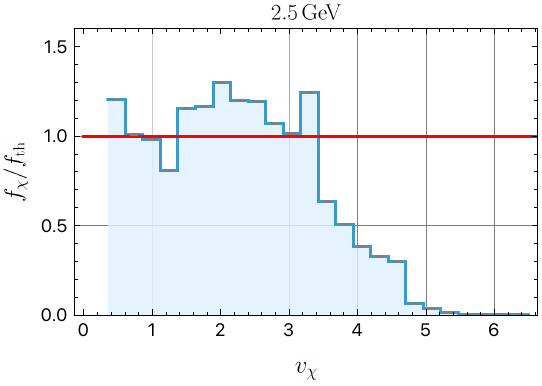} &
\includegraphics[width=0.485\linewidth]{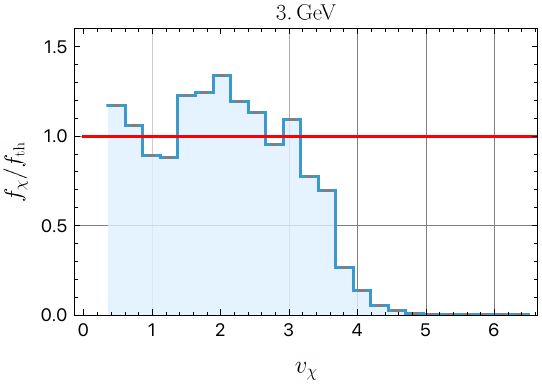} \\[1pt]
\includegraphics[width=0.485\linewidth]{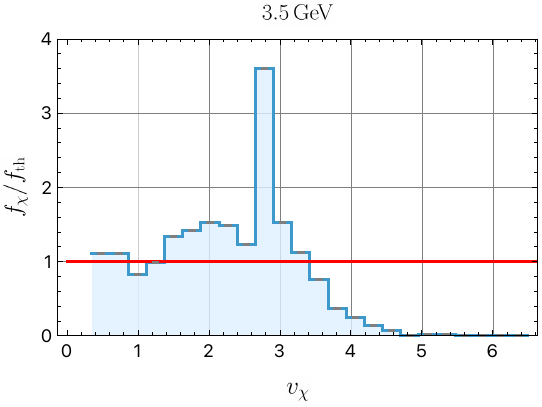} &
\includegraphics[width=0.485\linewidth]{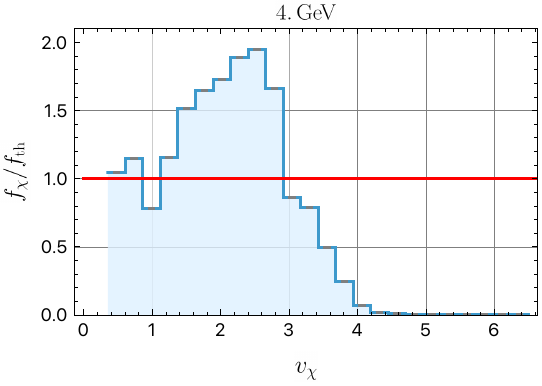}
\end{tabular}

\smallskip
{\small (b) $\beta=3$.}
\end{minipage}
\caption{Same quantity as in Fig.~\ref{fig:ratio-betas01}, now shown for $\beta=2$ in group (a) and $\beta=3$ in group (b), each over the mass range $m_\chi=0.5$--$4.0~\mathrm{GeV}$ in steps of $0.5~\mathrm{GeV}$.}
\label{fig:ratio-betas23}
\end{figure*}

Figures~\ref{fig:ratio-betas01} and~\ref{fig:ratio-betas23} show the ratio $f_\chi/f_{\rm th}$ for several benchmark masses and increasing values of the barrier strength $\beta$.
The ratio is most consequential near the high-velocity end, since evaporation is determined by the population closest to the escape condition rather than by the bulk of the distribution.

In the gravity-only case $\beta=0$, deviations from the isothermal reference follow the same broad pattern across the benchmark masses.
The ratio $f_\chi/f_{\rm th}$ shows a mild excess at very low velocities, falls below unity through the bulk of the distribution, and becomes increasingly depleted as $v$ approaches the near-escape region~\cite{Busoni:2017mhe}.
Although the distribution is exponentially suppressed close to the escape boundary, the asymptotic high-velocity tail determines the evaporation rate.
Its mass-dependent deformation is therefore essential for a controlled determination of the evaporation rate~\cite{Gould:1987ju,Liang:2016yjf,Busoni:2013kaa}.
For some benchmark masses, and more prominently once the barrier is present, the ratio can also exceed unity at low velocities or over an intermediate velocity interval.
Because $T_\chi$ is fixed by the equilibrium energy-exchange condition, a single-temperature isothermal reference matches an averaged energy scale yet does not reproduce the full shape of the non-thermal distribution~\cite{Widmark:2017yvd,Garani:2021feo}.

For $\beta>0$, we include an additional in-medium attraction arising from a light real scalar mediator.
In the non-relativistic regime, when the mediator range exceeds microscopic length scales, the solar medium sources a smooth mean-field potential whose magnitude is set by the local SM number density.
Because the density profile is strongly core-peaked, this attraction deepens the potential predominantly at small radii and leaves the exterior kinematics close to those in the gravity-only case.
Relative to $\beta=0$, a particle that scatters in the interior must overcome a larger energy deficit to escape, and the kinematic condition for evaporation becomes more sensitive to the radius of the last collision, providing the kinematic origin of an evaporation barrier.

This core-localized modification leaves a characteristic signature in the projected velocity distribution.
In the gravity-only limit $\beta=0$, the ratio $f_\chi/f_{\rm th}$ varies monotonically with velocity, showing a mild enhancement at low velocities and a progressively stronger depletion toward the near-escape region~\cite{Gould:1987ju,Liang:2016yjf,Busoni:2017mhe}.
Once the in-medium contribution is included, Figs.~\ref{fig:ratio-betas01} and~\ref{fig:ratio-betas23} show a non-monotonic depletion-enhancement structure.
After exhibiting a depleted band over a finite interval at low or intermediate velocities, the ratio rises above unity over a bound interval at larger values of $v$ before the final suppression close to the cutoff.
In several benchmark cases, this depletion-enhancement sequence appears more than once.

The in-medium attraction reshapes the bound region and the escape energetics in the solar interior~\cite{Acevedo:2023owd}. In the orbit-space formulation, the resulting depletion-enhancement structure reflects how occupation is redistributed in equilibrium among orbit families.
A typical bound orbit spends most of its time at large radii, and this long residence time dominates the very low-velocity weight.
Energy exchange is dominated by brief passages through the hot, dense core, which populate the large-$v$ region of the distribution.
The barrier primarily modifies the interior kinematics and biases the non-thermal distribution toward core-crossing trajectories, shifting weight toward the large-$v$ region of the spectrum relative to the isothermal reference over a finite range.
Meanwhile, weakly bound trajectories continue to support the low-velocity excess.
The intermediate-velocity band is supplied mainly by trajectories with perihelia in the transition region between the barrier-modified interior and the gravity-dominated exterior.
After the bound $(E,L)$ domain is compressed and reorganized, this interpolating set is not replenished efficiently, leaving a depleted intermediate-velocity interval upon projection onto velocity space~\cite{Liang:2016yjf,Widmark:2017yvd}. By contrast, the preferential retention and repeated processing of core-crossing trajectories shift additional weight into a bound interval at larger values of $v$, producing the subsequent enhancement.

\FloatBarrier
\section{Evaporation rates and evaporation mass}\label{evap-threshold}

In this section we assemble the physical ingredients governing the time evolution of the solar-captured dark-matter population, including capture, evaporation, and annihilation. Our analysis is based on the standard local scattering formalism for DM interactions in a celestial body, originally developed by Press--Spergel and Gould, which has become the foundation of modern treatments of solar DM capture and evaporation~\cite{Press:1985ug,Gould:1987ir,Gould:1987ju,Gould:1989tu}.
Unless stated otherwise, all collision rates in this section are evaluated for the same spin-independent interaction used throughout this work, namely the non-relativistic operator $\hat{\mathcal O}_1=\mathbf{1}$~\cite{Fan:2010gt,Fitzpatrick:2012ix,Anand:2013yka}. The microscopic interaction strength is therefore fully specified by the reference DM--proton cross section $\sigma_p$, and the differential cross section reduces to the standard isotropic contact form employed in Ref.~\cite{Liang:2016yjf}. Accordingly, the barrier parameter $\beta$ enters the collision terms only indirectly through the modified total potential $\phi_{\rm tot}$ and the corresponding escape-velocity profile, not through an explicit momentum-dependent modification of $d\sigma_{\chi A}/dv$.

We calculate the differential rate \( R_A(w\!\to\!v) \) at which a DM particle with velocity \( w \)
scatters to a final velocity \( v \), defined as
\begin{equation}
R_A(w \!\to\! v)
= n_A\!\int\! d^3u_A\, f_A(\mathbf{u}_A)\,
\frac{d\sigma_{\chi A}(|\mathbf{w}-\mathbf{u}_A|)}{dv}\,|\mathbf{w}-\mathbf{u}_A|\,,
\label{eq:RA_def}
\end{equation}
where $d\sigma_{\chi A}/dv$ is the differential cross section for DM--nucleus scattering, evaluated as a function of the relative speed $|\mathbf{w}-\mathbf{u}_A|$.
For the isospin-conserving $\hat{\mathcal O}_1$ interaction with equal proton and neutron couplings, we define the reference DM--proton cross section as
\begin{equation}
\sigma_p \equiv \frac{c_1^2\,\mu_p^2}{\pi}\,,
\label{eq:sigmap_def}
\end{equation}
where $c_1$ is the nucleon coupling of $\hat{\mathcal O}_1$ and $\mu_p$ is the DM--proton reduced mass. The corresponding zero-momentum spin-independent DM--nucleus cross section is then
\begin{equation}
\sigma_{\chi A}^{(0)}=\sigma_p\,\frac{\mu_A^2}{\mu_p^2}\,A^2\,,
\label{eq:sigmaA0}
\end{equation}
with $\mu_A$ the DM--nucleus reduced mass and $A$ the nuclear mass number.
Here $n_A(r)$ is the local number density of nuclear species $A$, and $\mathbf{u}_A$ is drawn from the local Maxwellian distribution in the solar rest frame.
The quantity $R_A(w\!\to\!v)$ is the local differential scattering rate per unit final speed and serves as the microscopic transition rate entering capture and evaporation.

The velocity distribution of solar nuclei follows the Maxwellian form,
\begin{equation}
f_A(\mathbf{u}_A)
= \left( \frac{m_A}{2\pi T_\odot} \right)^{3/2}
\exp\!\left[-\,\frac{m_A u_A^2}{2T_\odot}\right],
\label{eq:fA_Maxwell}
\end{equation}
where $T_\odot$ is the local solar temperature.
Throughout this work, the background profiles $n_A(r)$ and $T_\odot(r)$ are taken from a Standard Solar Model
with the GS98 composition, and the local escape velocity $v_{\rm esc}(r)$ is calculated consistently with the adopted
background potential, namely $\phi_{\rm grav}$ for $\beta=0$ and $\phi_{\rm tot}=\phi_{\rm grav}+\phi_{\rm barrier}$ for the barrier benchmarks~\cite{Grevesse:1998bj,Serenelli:2009yc}.

Integrating over a chosen final-velocity domain gives the corresponding local scattering rate per DM particle. The two choices used below describe evaporation and capture, respectively, and are written as
\begin{align}
\Omega^{+}(w\,|\,v_{\rm esc}) &= \sum_A \int_{v_{\rm esc}}^{\infty} R_A(w\!\to\! v)\,dv, \label{eq:Omega_plus}\\
\Omega^{-}(w\,|\,v_{\rm esc}) &= \sum_A \int_{0}^{v_{\rm esc}} R_A(w\!\to\! v)\,dv. \label{eq:Omega_minus}
\end{align}
Here, $\Omega^{\pm}(w|v_{\rm esc})$ depend on the radial coordinate $r$ through the density and velocity distributions of solar nuclei and the escape velocity $v_{\rm esc}$.
The separation at $v_{\rm esc}$ makes explicit the physical distinction between down-scattering into bound phase space (capture) and up-scattering into the unbound region (evaporation).

The capture rate is obtained by convolving the down-scattering probability with the incident halo flux.
A DM particle entering the Sun with asymptotic velocity $u$ has a local speed
$w \equiv \sqrt{u^{2} + v_{\rm esc}^{2}}$ at radius $r$.
Using the solar-frame halo velocity distribution $f(u)$, we obtain
\begin{equation}
C_{\odot}\;=\;\int_0^{R_\odot}\!4\pi r^2\,dr\int_0^\infty\!du\,\frac{f(u)}{u}\,w\;\Omega^{-}\!\left(w\,|\,v_{\rm esc}\right)\,.
\label{eq:Codot_R}
\end{equation}
This expression is evaluated in either the purely gravitational potential or the combined potential $\phi_{\rm tot}=\phi_{\rm grav}+\phi_{\rm barrier}$, characterized by the central barrier strength $\beta$.
Because the local escape velocity $v_{\rm esc}$ is fixed by the adopted potential, the presence of $\phi_{\rm barrier}$ reshapes the kinematic boundary separating gravitationally confined trajectories from unbound ones, and consequently affects both capture and evaporation through the integration limits in Eqs.~\eqref{eq:Omega_plus} and~\eqref{eq:Omega_minus}.

For later use, we evaluate $C_\odot$ numerically for each benchmark potential and provide compact empirical fits as functions of $x\equiv m_\chi/\mathrm{GeV}$. The fits cover DM masses from $0.1$ to $4~\mathrm{GeV}$ for $\beta=1,2,3$, while for the gravity-only case ($\beta=0$) the valid range is from $1.5$ to $4~\mathrm{GeV}$.
In evaluating the capture rates, we adopt the standard halo-model benchmark of an isothermal DM halo with local density $\rho_\chi = 0.3~\mathrm{GeV}\,\mathrm{cm}^{-3}$ and a Maxwellian velocity distribution with dispersion $v_0 = 220~\mathrm{km}\,\mathrm{s}^{-1}$, truncated at the Galactic escape velocity $544~\mathrm{km}\,\mathrm{s}^{-1}$~\cite{Baudis:2012review}.
Using the dimensionless mass variable $x \equiv m_\chi/\mathrm{GeV}$, the fitted polynomial expressions are

\begin{align}
C_{\odot}^{\beta=0}
&\simeq
\Bigl(1.33419-1.00356x+0.689278x^2-0.264090x^3 \nonumber\\
&\hspace{2.5em}
+\,0.0564136x^4-0.00618640x^5+0.000263822x^6\Bigr) \nonumber\\
&\hspace{2.5em}
\times \left(\frac{\sigma_{p}}{10^{-40}\,\mathrm{cm^2}}\right)
\times 10^{26}\,\mathrm{s^{-1}},
\label{eq:fitC_beta0}\\[3pt]
C_{\odot}^{\beta=1}
&\simeq
\Bigl(4.39567-11.4565x+15.6367x^2-10.7407x^3 \nonumber\\
&\hspace{2.5em}
+\,3.88566x^4-0.705974x^5+0.0507044x^6\Bigr) \nonumber\\
&\hspace{2.5em}
\times \left(\frac{\sigma_{p}}{10^{-40}\,\mathrm{cm^2}}\right)
\times 10^{26}\,\mathrm{s^{-1}},
\label{eq:fitC_beta1}\\[3pt]
C_{\odot}^{\beta=2}
&\simeq
\Bigl(6.05922-16.4128x+22.5964x^2-15.5922x^3 \nonumber\\
&\hspace{2.5em}
+\,5.65478x^4-1.02877x^5+0.0739363x^6\Bigr) \nonumber\\
&\hspace{2.5em}
\times \left(\frac{\sigma_{p}}{10^{-40}\,\mathrm{cm^2}}\right)
\times 10^{26}\,\mathrm{s^{-1}},
\label{eq:fitC_beta2}\\[3pt]
C_{\odot}^{\beta=3}
&\simeq
\Bigl(7.77909-21.667x+30.1129x^2-20.9274x^3 \nonumber\\
&\hspace{2.5em}
+\,7.63277x^4-1.39496x^5+0.100625x^6\Bigr) \nonumber\\
&\hspace{2.5em}
\times \left(\frac{\sigma_{p}}{10^{-40}\,\mathrm{cm^2}}\right)
\times 10^{26}\,\mathrm{s^{-1}}.
\label{eq:fitC_beta3}
\end{align}

The evaporation rate is obtained by integrating the up-scattering probability over the non-thermal distribution of dark matter $f_\chi(r,w)$, yielding
\begin{equation}
E_{\odot}\;=\;\int dr\int dw\, f_\chi(r,w)\;\Omega^{+}\!\left(w\,|\,v_{\rm esc}\right)\,.
\label{eq:Eodot_R}
\end{equation}
Here the non-thermal distribution is essential for evaporation.
For each potential strength $\beta$, we fit the equilibrium results over the relevant DM mass ranges, from $1.5$ to $4~\mathrm{GeV}$ for $\beta=0$ and from $0.1$ to $4~\mathrm{GeV}$ for $\beta=1,2,3$.
The corresponding empirical fits are
\begin{align}
E_{\odot}^{\beta=0}
&\simeq
8.78696\times10^{-6}\,
\exp\!\left[-2.08188\,(x^{1.23}+x^{-0.03})\right] \nonumber\\
&\hspace{2.5em}
\times \left(\frac{\sigma_{p}}{10^{-40}\,\mathrm{cm^2}}\right)\,\mathrm{s^{-1}},
\label{eq:fitE_beta0}
\\[3pt]
E_{\odot}^{\beta=1}
&\simeq
4.93563\times10^{-5}\,
\exp\!\left[-3.69456\,(x^{1.33}+x^{-0.03})\right] \nonumber\\
&\hspace{2.5em}
\times \left(\frac{\sigma_{p}}{10^{-40}\,\mathrm{cm^2}}\right)\,\mathrm{s^{-1}},
\label{eq:fitE_beta1}
\\[3pt]
E_{\odot}^{\beta=2}
&\simeq
2.23635\times10^{-4}\,
\exp\!\left[-5.35771\,(x^{1.39}+x^{-0.03})\right] \nonumber\\
&\hspace{2.5em}
\times \left(\frac{\sigma_{p}}{10^{-40}\,\mathrm{cm^2}}\right)\,\mathrm{s^{-1}},
\label{eq:fitE_beta2}
\\[3pt]
E_{\odot}^{\beta=3}
&\simeq
4.35859\times10^{-6}\,
\exp\!\left[-5.94803\,(x^{1.53}+x^{-0.03})\right] \nonumber\\
&\hspace{2.5em}
\times \left(\frac{\sigma_{p}}{10^{-40}\,\mathrm{cm^2}}\right)\,\mathrm{s^{-1}}.
\label{eq:fitE_beta3}
\end{align}

The radial number-density profile entering the annihilation term is obtained from the local distribution by
\begin{equation}
n_\chi(r)=\frac{N_\chi}{4\pi r^2}\int_0^{v_{\rm esc}(r)}\!dv\, f_\chi(r,v)\,,
\label{eq:nchi_from_frv}
\end{equation}
which satisfies $\int n_\chi(r)\,d^3r=N_\chi$ when the radial integral is taken over the full bound orbital support. In the solar scattering integrals above, only the solar-interior region contributes because the target densities vanish outside $R_\odot$. The annihilation coefficient is then defined in terms of the effective annihilation volume that characterizes the spatial overlap of the dark-matter distribution~\cite{Busoni:2013kaa,Garani:2017jcj},
\begin{equation}
A_{\odot}=\frac{\langle\sigma v\rangle_{\odot}}{V_{\rm eff}},\qquad
V_{\rm eff}\equiv\frac{\left[\int n_\chi(r)\,d^3r\right]^2}{\int n_\chi^2(r)\,d^3r},
\label{eq:Aodot_veff_def}
\end{equation}
which is a convenient parametrization of the annihilation term appearing in the standard evolution equation
for the captured population.
The effective annihilation volumes are well captured by the following empirical fits,
\begin{align}
V_{\rm eff}^{\beta=0}
&\simeq
7.21970\times10^{29}
\left(\frac{4~\mathrm{GeV}}{m_\chi}\right)^{2.23889}\ \mathrm{cm^3},
\label{eq:fitVeff_beta0}
\\[3pt]
V_{\rm eff}^{\beta=1}
&\simeq
1.77388\times10^{28}
\left(\frac{4~\mathrm{GeV}}{m_\chi}\right)^{2.62957}\ \mathrm{cm^3},
\label{eq:fitVeff_beta1}
\\[3pt]
V_{\rm eff}^{\beta=2}
&\simeq
6.30156\times10^{27}
\left(\frac{4~\mathrm{GeV}}{m_\chi}\right)^{2.5121}\ \mathrm{cm^3},
\label{eq:fitVeff_beta2}
\\[3pt]
V_{\rm eff}^{\beta=3}
&\simeq
4.28607\times10^{27}
\left(\frac{4~\mathrm{GeV}}{m_\chi}\right)^{2.32788}\ \mathrm{cm^3}.
\label{eq:fitVeff_beta3}
\end{align}

With $C_\odot$, $E_\odot$, and $A_\odot$ specified, the total number of captured DM particles obeys the standard population equation~\cite{Busoni:2013kaa,Garani:2017jcj}
\begin{equation}
\frac{dN_\chi}{dt} = C_\odot - E_\odot N_\chi - A_\odot N_\chi^2,
\label{eq:master}
\end{equation}
and the corresponding DM annihilation rate in the Sun is
\begin{equation}
\Gamma_A=\frac{1}{2}A_\odot N_\chi^2.
\label{eq:GammaA}
\end{equation}
For time-independent coefficients and the initial condition $N_\chi(0)=0$, the solution can be written as~\cite{Busoni:2013kaa,Garani:2017jcj}
\begin{equation}
N_\chi(t)=
\frac{C_\odot \tanh\!\left(t/\tau_e\right)}
{\tau_e^{-1}+\frac{1}{2}E_\odot\tanh\!\left(t/\tau_e\right)}\,,
\label{eq:Nchi_solution}
\end{equation}
with the equilibration timescale
\begin{equation}
\tau_e \equiv \left(C_\odot A_\odot + \frac{E_\odot^2}{4}\right)^{-1/2}.
\label{eq:taue}
\end{equation}
This timescale controls the approach to equilibrium and is used below to discuss the regime maps.

These fitting formulae are intended as phenomenological interpolants over the mass ranges quoted above. We conservatively assign a $10\%$--$20\%$ interpolation uncertainty to the quoted $(C_\odot,E_\odot,V_{\rm eff})$ fits within their stated validity windows, with the larger deviations expected near the endpoints of the fitted mass ranges. They should therefore be used for efficient interpolation rather than precision extrapolation beyond the quoted domains.

\begin{figure*}[!t]
  \centering
  \setlength{\tabcolsep}{3pt}
  \renewcommand{\arraystretch}{1.0}
  \begin{tabular}{cc}
    \includegraphics[width=0.475\textwidth]{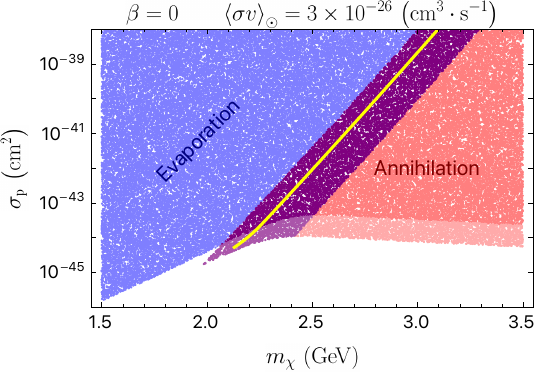} &
    \includegraphics[width=0.475\textwidth]{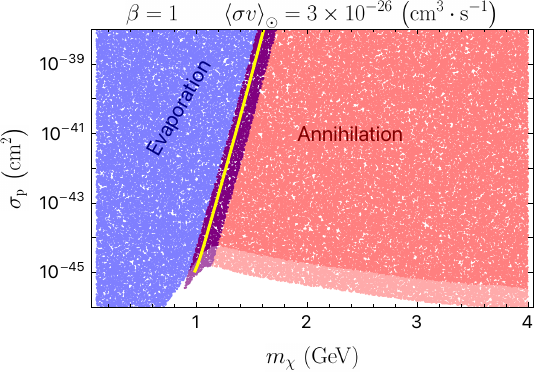} \\[4pt]
    \includegraphics[width=0.475\textwidth]{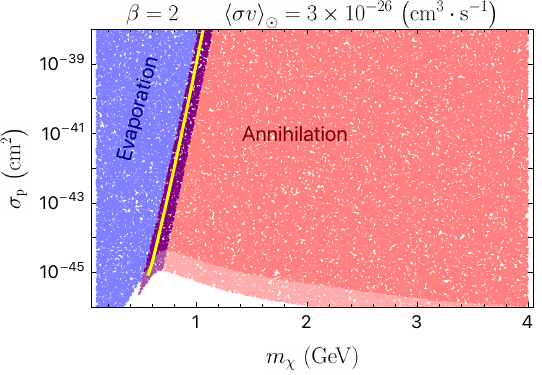} &
    \includegraphics[width=0.475\textwidth]{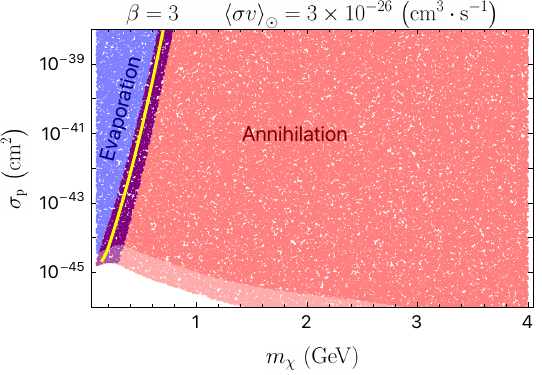}
  \end{tabular}
  \caption{Regime structure in the $(m_\chi,\sigma_p)$ plane for $\beta=0,1,2,3$, shown from top left to bottom right. The darker-colored regions mark near-equilibrium parameter space, $\tanh(t_{\odot}/\tau_e)\simeq 1$, while the lighter regions correspond to $0.9\le \tanh(t_{\odot}/\tau_e)\le 1$ for reference. In the blue (red) region, annihilation (evaporation) plays a subdominant role in the evolution of the solar dark-matter number. The purple belt marks the transition zone in which evaporation and annihilation are comparably important. The yellow curve traces the evaporation mass $m_{\rm evap}$.}
  \label{fig:Nequilibrium}
\end{figure*}

Using the criterion introduced in Refs.~\cite{Busoni:2013kaa,Garani:2017jcj}, we operationally define the evaporation mass $m_{\rm evap}$ as the minimum testable dark-matter mass, namely the threshold at which the deviation from the evaporation--capture equilibrium value reaches ten percent of the actual solar population,
\begin{equation}
\left|\,N_\chi-\frac{C_{\odot}}{E_{\odot}}\,\right| = 0.1\,N_\chi\,.
\label{eq:mevap_def}
\end{equation}
We adopt this definition of $m_{\rm evap}$ throughout this work. Below this mass scale, evaporation is sufficiently efficient to substantially suppress the accumulation of dark matter in the Sun.

We classify the parameter space by comparing the evaporation and annihilation timescales through the dimensionless ratio $E_{\odot}^{2}/(4C_{\odot}A_{\odot})$.
Regions with $E_{\odot}^{2}/(4C_{\odot}A_{\odot}) \le 0.1$ are annihilation-dominated. Regions with $E_{\odot}^{2}/(4C_{\odot}A_{\odot}) \ge 10$ are evaporation-dominated. The intermediate range marks the transition band where evaporation and annihilation are comparably important.
Based on these criteria and assuming the canonical thermal $s$-wave annihilation cross section $\langle\sigma v\rangle_\odot = 3\times10^{-26}\,\mathrm{cm^{3}\cdot s^{-1}}$~\cite{Jungman:1995df,Steigman:2012relic}, we construct the regime maps in Fig.~\ref{fig:Nequilibrium}, illustrating the boundaries among the three regimes for representative benchmarks and indicating the evaporation mass $m_{\rm evap}$ as the yellow curve.

\section{Discussion and Conclusions}\label{discussion}

An in-medium evaporation barrier makes the weakly bound population near the escape boundary a quantitatively important component of the solar-captured dark-matter distribution. Because evaporation is exponentially sensitive to the phase-space density in this near-threshold region, reliable predictions require the actual non-thermal distribution to be resolved rather than replaced by a single-temperature prescription calibrated only to the bulk population. Our orbit-space treatment addresses this point directly by determining the distribution over bound trajectories labeled by specific energy and angular momentum.

Relative to the gravity-only case, the barrier modifies the capture and escape kinematics by deepening the potential well and raising the local escape velocity. This increases the capture rate and suppresses evaporation at fixed microphysics. The non-thermal distribution also shifts toward more tightly bound states and smaller angular-momentum values, consistent with scattering being dominated by the hot, dense solar core. That redistribution reshapes both the near-threshold tail and the spatial concentration of the bound population. As a consequence, the control ratio $E_\odot^{2}/(4C_\odot A_\odot)$ decreases and the evaporation mass $m_{\rm evap}$ shifts to smaller values. The same phase-space restructuring also leaves a distinctive imprint on the projected velocity distribution, producing a non-monotonic depletion-enhancement pattern with a depleted interval followed by an enhancement at larger bound values of $v$ before the final suppression near the cutoff.

The intermediate-velocity depletion can be understood from how different orbit families populate the local spectrum once the bound domain is deformed by the barrier. The low-velocity weight is dominated by long residence times at large radii, whereas diffusion in energy and angular momentum is driven by brief passages through the core. The barrier preferentially retains and reprocesses core-crossing trajectories, while the orbit families that interpolate between outer-dominated and core-dominated contributions become less efficiently populated in equilibrium. Upon projection, this produces a depleted intermediate-velocity interval together with an enhanced interval at larger values of $v$. In some benchmark cases the same mechanism can operate across multiple orbital bands, yielding multiple separated depletion-enhancement features.

Taken together, these results show that barrier-induced reshaping of the near-threshold tail can directly modify the solar dark-matter population and the associated neutrino signal within the present fixed-background framework. In the barrier regime, the near-threshold tail is therefore not a minor correction but a controlling ingredient in reliable evaporation estimates. For phenomenological applications, the fitted $(C_\odot, E_\odot, V_{\rm eff})$ functions and the regime maps in the $(m_\chi,\sigma_p)$ plane provide practical tools for incorporating these effects. Several extensions are well motivated. It will be important to relate the phenomenological barrier strength $\beta$ more directly to mediator microphysics and to quantify finite-range effects beyond the simplest local-density scaling. Additional scattering channels, including electron scattering and more general momentum- or velocity-dependent operators, can further modify the near-threshold tail and deserve a systematic treatment. Extending the gravity-only computation to masses well below the evaporation mass will likewise require improved numerical control in the weakly bound sector of the discretized orbit-space domain, where evaporation rapidly depletes near-threshold trajectories.

The present analysis also has several clear limits of validity. First, capture, thermalization, and evaporation are all evaluated with the spin-independent $\hat{\mathcal O}_1$ interaction, while the light mediator is included only through the in-medium potential. This factorized treatment is appropriate only when mediator-induced binary scattering remains subdominant. If the mediator contribution to binary scattering becomes appreciable during capture or thermalization, the full momentum-dependent differential cross section should be included. Second, the mediator contribution is treated through the local-density form of the in-medium potential, whereas finite-range corrections can modify the detailed barrier profile once the mediator range becomes comparable to solar-structure scales. Third, in a minimal scalar Yukawa realization the captured DM would itself source an additional mean field and backreact on the total potential. That DM-sourced contribution is not included here, because the present work is restricted to the fixed-background problem of a prescribed SM-sourced barrier. A complete treatment would require solving simultaneously for the mediator profile and the corresponding DM non-thermal distribution in the resulting total potential. Fourth, the gravity-only computation at very low masses remains numerically difficult because rapid evaporation depletes the near-threshold orbit-space bins adjacent to the escape boundary. Extending that regime will therefore require improved control of the near-threshold tail.

\section*{Acknowledgements}
The author is especially grateful to Prof.~Zhengliang Liang for many helpful conversations and to Prof.~Chao-Qiang Geng for helpful advice. The author wishes to acknowledge Prof.~Kenji Kadota for past guidance and inspiration. This work was supported by the School of Fundamental Physics and Mathematical Sciences, Hangzhou Institute for Advanced Study, University of Chinese Academy of Sciences (UCAS) and by the Chinese Academy of Sciences.
\bibliographystyle{elsarticle-num}
\bibliography{example}
\end{document}